\documentstyle[prl,manuscript,aps]{revtex}
\begin{document}
\draft
\title{A proposal to search for a monochromatic component of\\
solar axions using $^{57}$Fe}
\author{Shigetaka Moriyama
\thanks{E-mail address: moriyama@icepp.s.u-tokyo.ac.jp}}
\address{Department of Physics, School of Science,
University of Tokyo,\\
7-3-1 Hongo, Bunkyo-ku, Tokyo 113, Japan}
\date{April 18, 1995}
\maketitle
\begin{abstract}
A new experimental scheme is proposed
to search for almost monochromatic solar axions,
whose existence has not been discussed heretofore.
The axions would be produced
when thermally excited $^{57}$Fe in the sun
relaxes to its ground state and could be detected
via resonant excitation of the same nuclide in a laboratory.
A detailed calculation shows that
the rate of the excitation is up to
order 10 events/day/kg-$^{57}$Fe.
The excitation can be detected efficiently using bolometers
or liquid scintillators.
\end{abstract}
\pacs{}

The most attractive solution of the strong $CP$ problem
is to introduce the Peccei-Quinn global symmetry
which is spontaneously broken at energy scale $f_a$\cite{PQ77}.
The original axion model assumed that $f_a$ is equal to
the electroweak scale.
Although it has been experimentally excluded,
variant ``invisible'' axion models
are still viable, in which $f_a$ is assumed to be very large;
since coupling constants of the axion with matter are
inversely proportional to $f_a$, experimental detection
becomes very difficult.
Such models are referred to as hadronic\cite{Kim}
and Dine-Fischler-Srednicki-Zhitnitski\v\i (DFSZ)\cite{DFSZ} axions.
At present, these ``invisible'' axions are constrained by
laboratory searches and by astrophysical and cosmological arguments.
One frequently quoted window for $f_a$
which escapes all the phenomenological
constraints is $10^{10}$--$10^{12}$\,GeV.
Besides this, there is another window
around $10^6$\,GeV for the hadronic axions which
have vanishing tree level coupling to the electron.
This is usually called the hadronic axion window.
Recently, a careful study\cite{Chang93} of the hadronic axion window
revealed that $f_a$ in the range $3\times 10^5$--$3\times 10^6$\,GeV
cannot be excluded by the existing arguments,
because most of them were based on the axion-photon coupling
which is the least known parameter
among those describing the low energy dynamics of the hadronic axions.

Although several authors\cite{Sikivie83,Bibber91,Paschos93}
proposed experimental methods to search for the axions
with $f_a$ around $10^6$\,GeV,
all of the methods are clearly based on the axion-photon coupling
both at the source and at the detector.
The methods utilizes only the Primakoff effect;
photons in the sun are converted into axions,
which are commonly called the solar axions,
and they are re-converted into x rays in a laboratory.
Thus there have been no experimental alternatives
to test the hadronic axion window
independently of the axion-photon coupling.

Owing to axion coupling to nucleons,
there is another component of solar axions.
If some nuclides in the sun have $M1$ transitions
and are excited thermally,
axion emission from nuclear deexcitation could be also possible.
$^{57}$Fe can be a suitable axion emitter
by the following reasons:
(i) $^{57}$Fe has an $M1$ transition
between the first excited state and the ground state,
(ii) the first excitation energy of $^{57}$Fe
is 14.4\,keV, which is not too high compared with the temperature
in the center of the sun $(\sim 1.3\,\rm keV)$\cite{Bahcall92},
and (iii) $^{57}$Fe is one of the stable isotopes of iron
(natural abundance 2.2\%),
which is exceptionally abundant
among heavy elements in the sun\cite{SolarInterior}.
If the axion exists, therefore,
strong emission of axions is expected
from this nuclide.

These monochromatic axions would excite the same nuclide
in a laboratory because the axions are Doppler broadened
due to thermal motion of the axion emitter in the sun
and thus some axions have energy suitable to excite the nuclide.

I propose to search for the axions by detecting this excitation.
Since both the emission and absorption occurs
via the axion-nucleon coupling
but not via the axion-photon coupling,
this method is free from the uncertainty
of the axion-photon coupling.
In addition, this method has merits that
there is no need to tune the detector to a mass of the axions
and that the mass can be large far beyond
that of the proposed experiment\cite{Bibber91},
in which it is restricted by high pressure of buffer gas.
In this letter, the detection rate of the resonant excitation
by the monochromatic solar axions is calculated
and experimental methods are discussed.

To estimate axion flux from the sun,
calculation can be performed as in Ref.\ \cite{Haxton91}.
The energy loss due to the axion emission is
\begin{eqnarray}
\delta E(T) &=&N \frac{2\exp(-\beta_T)}{1+2\exp(-\beta_T)}
\frac{1}{\tau_\gamma}
\frac{\Gamma_a}{\Gamma_\gamma}E_\gamma,
\label{eq:rate_per_1g}
\end{eqnarray}
where $N=2.9\times 10^{17}\rm g^{-1}$
is the number of $^{57}$Fe per 1\,g material
in the sun\cite{SolarInterior},
$\beta_T=(14.4\,{\rm keV})/kT$,
$\tau_\gamma=1.3\times 10^{-6}\rm\,sec$,
and $E_\gamma=14.4\rm\,keV$.
$\Gamma_a/\Gamma_\gamma$ represents
the branching ratio\cite{Haxton91},
\begin{eqnarray}
\frac{\Gamma_a}{\Gamma_{\gamma}} &=&
\frac{1}{2\pi\alpha}\frac{1}{1+\delta^2}
\left[
\frac{g_0\beta+g_3}{(\mu_0-1/2)\beta+\mu_3-\eta}
\right]^2,
\label{eq:bra_ratio}
\end{eqnarray}
where $\delta \sim 0$ is the $E2/M1$ mixing ratio.
$\mu_0$ and $\mu_3$ are
the isoscalar and isovector magnetic moments, respectively:
$\mu_0-1/2\sim 0.38$ and $\mu_3\sim 4.71$.
$\beta=-1.19$ and $\eta=0.80$ are
the nuclear-structure-dependent terms.
$g_0$ and $g_3$ are defined as\cite{Kaplan85}
\begin{eqnarray}
{\cal L} &=& a\bar{N}i\gamma_5(g_0+g_3\tau_3)N,\\
g_0&=&-7.8\times 10^{-8}
\left(\frac{6.2\times 10^6}{f_a/{\rm GeV}}\right)
\left(\frac{3F-D+2S}{3}\right),\\
g_3&=&-7.8\times 10^{-8}
\left(\frac{6.2\times 10^6}{f_a/{\rm GeV}}\right)
\left[(D+F)\frac{1-z}{1+z}\right],
\label{eq:g_3}\\
m_a&=&\frac{\sqrt{z}}{1+z}\frac{f_\pi m_\pi}{f_a}\nonumber\\
{}&=&1\,{\rm eV}\frac{\sqrt{z}}{1+z}
\frac{1.3\times 10^{7}}{f_a/{\rm GeV}},
\end{eqnarray}
where $D$ and $F$ denote the reduced matrix elements
for the $SU(3)$ octet axial vector currents
and $S$ characterizes the flavor singlet coupling.
The naive quark model (NQM) predicts $S=0.68$
\cite{Haxton91}, but measurements
\protect\cite{SMC} show that $S=0.22\pm0.1\pm0.1$.
$z=m_u/m_d\sim 0.56$ in the first order calculation.
$m_a$ is evaluated to be 1\,eV with
$z=0.56$ and $f_a=6.2\times 10^6\rm\,GeV$.
Using Eqs.\ (\ref{eq:bra_ratio})--(\ref{eq:g_3}),
Eq.\ (\ref{eq:rate_per_1g}) becomes
\begin{eqnarray}
\delta E(T)&=&4.6\times 10^4 {\,\rm erg\,g^{-1}\,s^{-1}}\nonumber\\
{}&&\times \left(\frac{10^6\,{\rm GeV}}{f_a}\right)^2
C^2 \exp(-\beta_T),\label{eq:rate_per_1g_DFSz}\\
C(D,F,S,z)&\equiv&-1.19\left(\frac{3F-D+2S}{3}\right)\nonumber\\
{}&&+(D+F)\frac{1-z}{1+z},
\label{eq:C}
\end{eqnarray}
where $\beta_T\gg 1$ is assumed in the solar interior.
Our estimation slightly differs from that of Ref.\ \cite{Haxton91}
because a different value of $^{57}$Fe abundance in the sun
is used\cite{SolarInterior}.

Eq.\ (\ref{eq:rate_per_1g_DFSz}) provides an estimation of
the differential axion flux at the earth,
\begin{eqnarray}
\frac{d\Phi(E_a)}{dE_a} &=& \frac{1}{4\pi R_{E}^2}
\int^{R_\odot}_0
\frac{1}{\sqrt{2\pi}\sigma(T)}
\exp\left[-\frac{(E_a-E_\gamma)^2}{2\sigma(T)^2}\right]\nonumber\\
{}&& \times
\frac{\delta E(T)}{E_\gamma} \rho(r) 4\pi r^2 dr,
\label{eq:spectrum}
\end{eqnarray}
where $R_E$ is the average distance
between the sun and the earth.
$R_\odot$ denotes the solar radius.
$T(r)$ and $\rho(r)$ are the temperature and
the mass density at the radius $r$, respectively.
$\sigma(T)=E_\gamma (kT/m)^{1/2}$ represents the Doppler broadening.
$m$ is the mass of the $^{57}$Fe nucleus.
It should be noted that the number of the iron atom
per unit mass is assumed to be uniform
as in the framework of the standard solar model (SSM)\cite{Bahcall92},
i.e. that $N$ is independent of $r$.
In addition, the SSM provides the mass density
and the temperature as a function of the radius $r$,
which is necessary for calculating Eq.\ (\ref{eq:spectrum}).
The values of the functions are taken
from Table XVI in Ref.\ \cite{Bahcall92}.
Thus Eq.\ (\ref{eq:spectrum}) can be evaluated
if one fixes $D,F,S,z,$ and $f_a$.
The sharp peak in Fig.\ \ref{fig:spectrum}
corresponds to the axion flux evaluated
with $D=0.77, F=0.48, S=0.68, z=0.56,$ and $f_a=10^6\,\rm GeV$.
Also shown is the expected axion flux
generated through the Primakoff effect\cite{Bibber91}.
It is a striking fact that substantial axion emission
is expected from the nuclear deexcitation.
The differential flux
at $E_\gamma$ is obtained to be
\begin{eqnarray}
A&=&2.0\times 10^{13}{\,\rm cm^{-2}\,s^{-1}\,keV^{-1}}
\left(\frac{10^6{\rm\,GeV}}{f_a}\right)^2C^2,
\label{eq:fluxatEgam}
\end{eqnarray}
where dependences on $D,F,S,$ and $z$ are included in $C$.
The effects of the nuclear recoil and
of the red shift due to the gravitation
of the sun are negligible.
The former decreases the axion energy
by only about $1.9\times 10^{-3}$\,eV
and the latter about $1.5\times 10^{-1}$\,eV,
which are negligibly small compared with the width of
the peak in Fig.\ \ref{fig:spectrum}.

In a laboratory, these axions would
resonantly excite $^{57}$Fe.
The rate of the excitation is calculated as follows.
It is a reasonable approximation
that $d\Phi(E_a)/dE_a=A$ over the natural width
of $^{57}$Fe, ${\cal O}(10\,\rm{neV})$,
around 14.4\,keV because the width of the peak
in Fig.\ \ref{fig:spectrum} is extremely broadened to about 5\,eV.
Hence the rate of the excitation per $^{57}$Fe nuclei is
\begin{eqnarray}
R_{N} &=& A \sigma_{0,a} \Gamma_{\rm tot}
\frac{\pi}{2},\label{eq:rateper1} \\
\sigma_{0,a}&=& 2\sigma_{0,\gamma}\frac{\Gamma_a}{\Gamma_\gamma},
\label{eq:sigma_a}
\end{eqnarray}
where $\sigma_{0,\gamma}=2.6\times 10^{-18}\rm\,cm^2$
is the maximum resonant cross section of
$\gamma$ rays\cite{DeRujula89} and
$\Gamma_{\rm tot}=4.7\times 10^{-12}\rm\,keV$ is the total decay width
of the first excited state of $^{57}$Fe.
The factor 2 in the Eq.\ (\ref{eq:sigma_a}) represents
the difference of the spin multiplicity between photons and axions.

Eq.\ (\ref{eq:fluxatEgam}) and Eq.\ (\ref{eq:rateper1}) now
allow us to calculate total detection rate
per unit mass of $^{57}$Fe in the laboratory,
\begin{eqnarray}
R &=& 3.0\times 10^2{\rm\,day^{-1}\,kg^{-1}}
\left(\frac{10^6\,{\rm\,GeV}}{f_a}\right)^4 C^4,
\end{eqnarray}
where $C$ depends on $D,F,S,$ and $z$ as shown in Eq.\ (\ref{eq:C}).
As for $D$ and $F$, measurements of
the nucleon and hyperon $\beta$ decays
show $D=0.77$ and $F=0.48$\cite{Haxton91}.
However, the estimations of $S$ and $z$ have
large uncertainties and ambiguity\cite{SMC,Kaplan86}.
In particular, $z$ might suffer large corrections
due to instanton effects\cite{Kaplan86} and
be significantly smaller than the value of
the first order calculation, 0.56.
Therefore, the detection rate should be represented
as a function of $S$ and $z$.
Fig.\ \ref{fig:rate} shows the contours
of the calculated detection rate with $f_a=10^6\,\rm GeV$.

We now turn to a discussion of experimental methods.
After the excitation of the nuclei by the axion,
the emission of a $\gamma$ ray, with an energy of 14.4\,keV, or
the emissions of internal conversion electron,
with an energy of 7.3\,keV,
and the subsequent atomic radiations will occur.
Since the attenuation length of the $\gamma$ ray is $20\,\mu$m
and the range of the electron is $0.2\,\mu$m in iron,
it is difficult to detect the $\gamma$ rays
or electrons outside iron.
However, this excitation is detected efficiently
(i) by a bolometric technique
with an absorber which contains $^{57}$Fe-enriched iron
or (ii) by using iron-loaded liquid scintillators.

Since the iron atoms have large magnetic moment,
magnons might appear in crystals containing iron atoms.
The additional contributions
(e.g. by the magnons of the ferrimagnetisms)
to the specific heat is unfavorable
because the essential point of the bolometric technique is
that the specific heat of the absorber is very small.
At least theoretically, however,
if the absorber is ferrimagnetic,
the specific heat due to the magnons of the ferrimagnetisms
can be reduced exponentially by applying strong magnetic field.
For example, therefore, crystals of ferrimagnetisms such as
FeO, $\gamma$-$\rm Fe_2O_3$, or $\rm Fe_3O_4$
with strong external magnetic field
can be used for the absorber.
The weak point of this technique is
that one must enrich $^{57}$Fe
because the total mass must be reduced.

The iron-loaded liquid scintillator
is a feasible detector to search for the monochromatic axions.
If one can load 10\% natural iron to the scintillator by weight,
the liquid scintillator with a total mass of 500\,kg
contain about 1-kg $^{57}$Fe.
The large mass is easily available,
but there might be large background.

The axion events are easily separated from background
because the axion events would form
a sharp peak in the energy spectrum.
Although $^{57}$Co contamination in the detector
might cause background events at the peak,
one can reject such events
using anti-coincidence with 122-keV $\gamma$ ray,
which is emitted in advance of 14.4-keV $\gamma$ ray.
It should be noted that a reanalysis of any existing data
cannot constrain the hadronic axion model
using the present calculation,
because the iron itself seems not to have been used
as a detector of radiation yet.

In summary, a new scheme to detect almost monochromatic
solar axions using resonant excitation of $^{57}$Fe is proposed.
$^{57}$Fe is rich in the sun and its first excitation energy
is low enough to be excited thermally.
Therefore, one can expect
the nuclear deexcitation accompanied with the axion emission.
Due to the Doppler effect associated
with the thermal motion of $^{57}$Fe in the sun,
a small portion of the axions from the nuclide
can be absorbed by the same nuclide in a laboratory.
The nuclide is considered as a well tuned detector of the axions.
A detailed calculation shows that the excitation rate
is up to order $10\,\rm day^{-1}kg^{-1}$.
Although it is difficult to detect the excitation outside the iron,
this excitation is detected efficiently by a bolometric technique
with an absorber which contains $^{57}$Fe-enriched iron
or by iron-loaded liquid scintillators.
I am planning an experiment to search for
the monochromatic axions from the sun in this new scheme.

I am very much indebted to Professor M.~Minowa
for his many suggestions to improve the manuscript and
for his helpful comments concerning experimental realities.
I wish to thank Y.~Inoue and
Y.~Kishimoto for discussions with them and
Y.~Inagaki, H.~Asakawa, and A.~Kawamura for their helpful comments.
Discussions with Y.~Ito, W.~Ootani, and K.~Nishigaki
helped me to understand the bolometric technique.

\begin{figure}
\caption{Differential flux of the axion from the sun.
The sharp peak corresponds to the axion emission
from the $^{57}$Fe deexcitation.
Broad part of the differential flux corresponds to
the axion generated through the Primakoff effect.}
\label{fig:spectrum}
\end{figure}
\begin{figure}
\caption{Contours of the detection rate as a function of $S$ and $z$.
The naive quark model (NQM) predicts $S=0.68$\protect\cite{Haxton91},
but measurements \protect\cite{SMC} show that $S=0.22\pm0.1\pm0.1$.}
\label{fig:rate}
\end{figure}

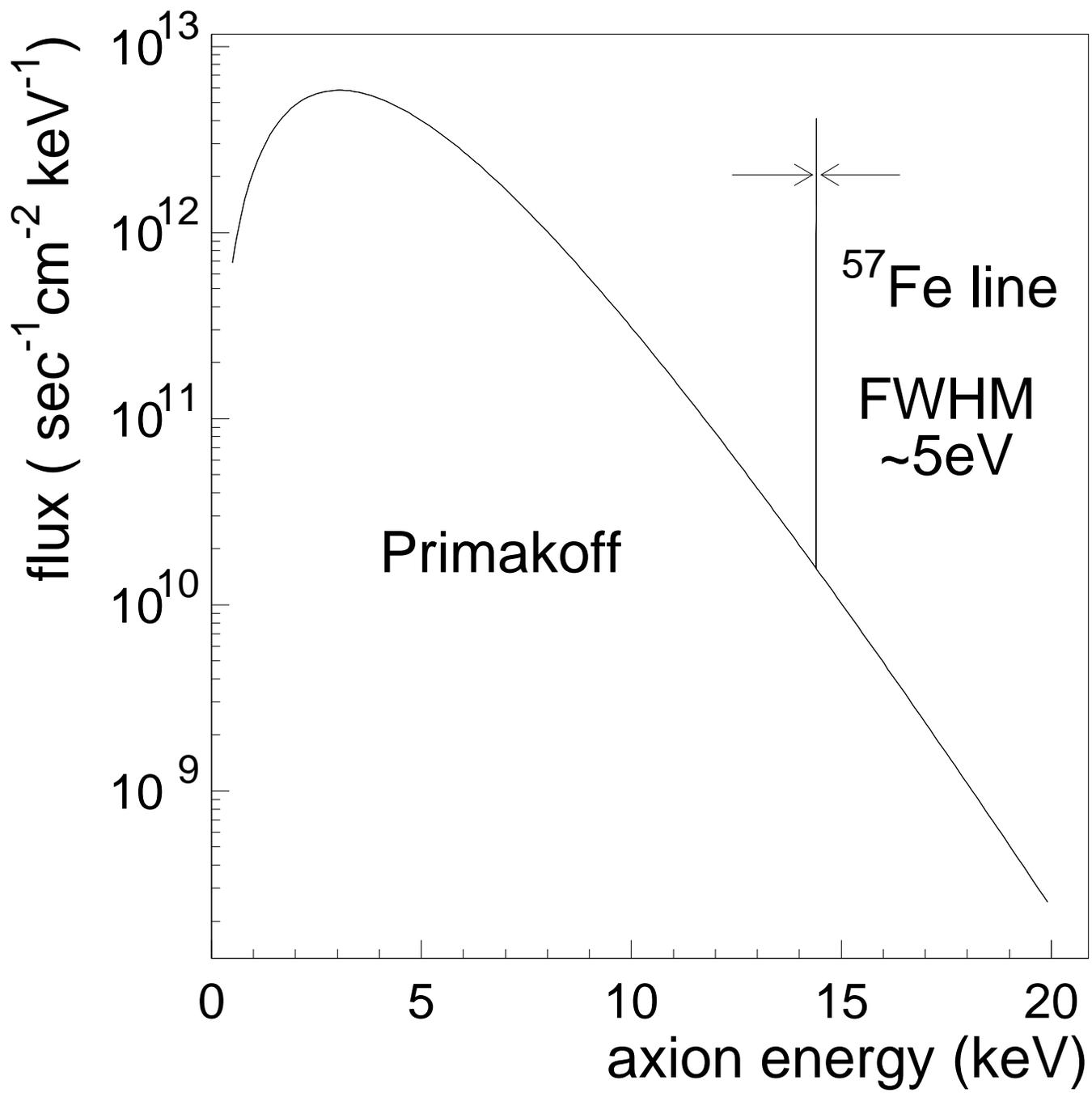
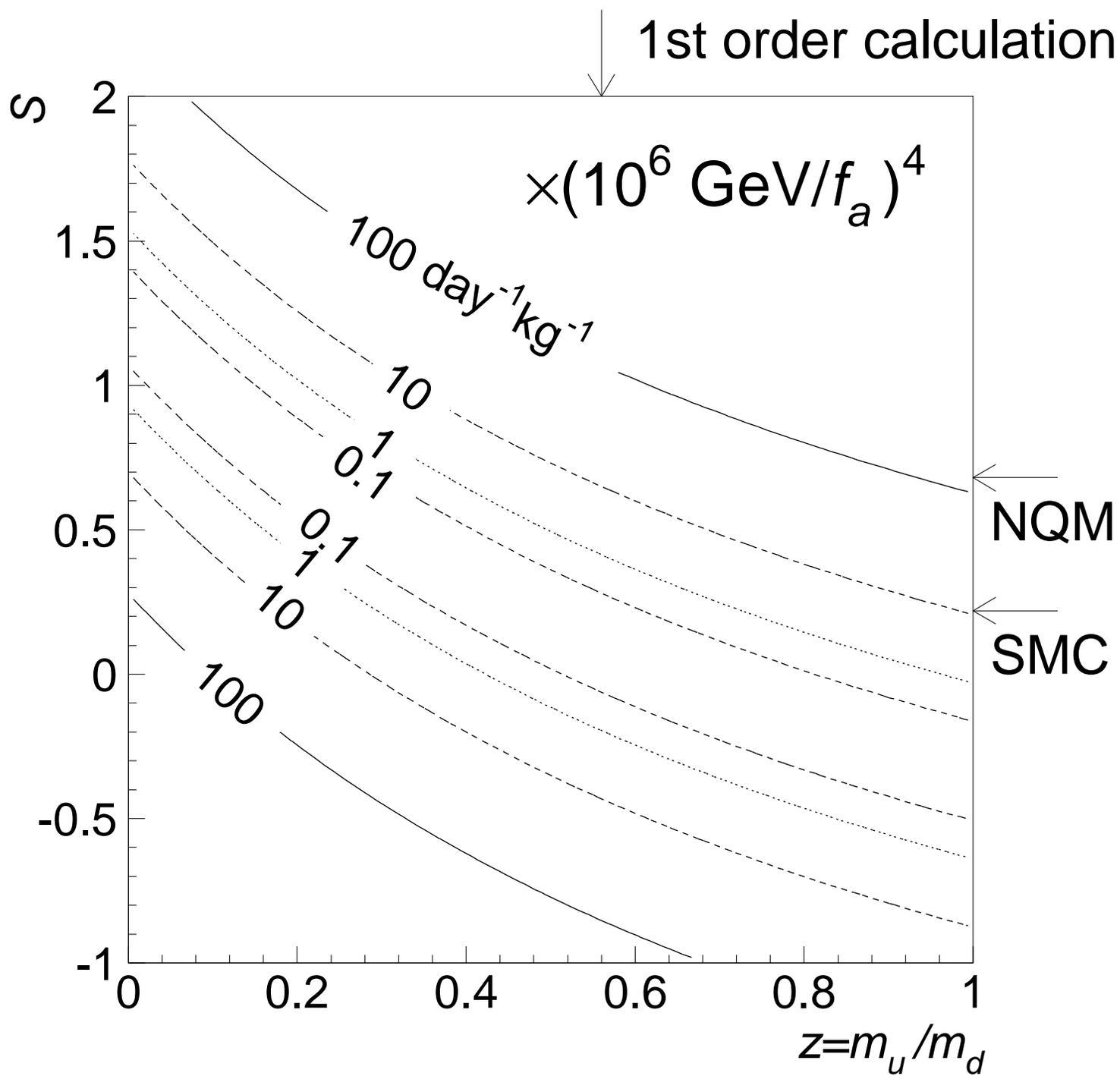
\end{document}